\documentclass[11pt,twoside]{article}

\usepackage{asp2010}
\resetcounters
\begin{document}

\title{Optical observations of PSR J1357$-$6429 field}

\author{Aida~Kirichenko,$^{1,2}$ Andrey~Danilenko,$^2$ 
Ronald~E.~Mennickent,$^3$ George~Pavlov,$^{1,4}$ Yury~Shibanov,$^{1,2}$
Sergey~Zharikov,$^5$ and Dmitry~Zyuzin$^{1,2}$
\affil{
$^1$St. Petersburg State Polytechnical Univ., Politekhnicheskaya 29, 
St. Petersburg, 195251, Russia\\
$^2$Ioffe Physical Technical Institute, Politekhnicheskaya 26, St. Petersburg, 194021, Russia\\
$^3$Department of Astronomy, Universidad de Concepcion, Casilla 160-C, Concepcion, Chile\\
$^4$Department of Astronomy \& Astrophysics, Pennsylvania State University, PA 16802, USA\\
$^5$Observatorio Astron\'{o}mico Nacional SPM, Instituto de Astronom\'{i}a, Universidad Nacional 
Aut\'{o}nomia de Mexico, Ensenada, BC, Mexico}
}

\begin{abstract}
PSR J1357$-$6429 is a Vela-like radio pulsar that has been recently detected with 
\textit{Chandra} and \textit{Fermi},
which, like Vela, powers a compact X-ray pulsar wind nebula and 
X-ray-radio plerion associated with an extended TeV source. 
We present our deep optical observations with the Very Large Telescope 
to search for an optical counterpart of the pulsar and its nebula. 
We detected a point-like source in $V$, $R$, and $I$ bands 
whose position is in agreement with the X-ray position of the pulsar, 
and whose colours are distinct from those of ordinary stars. 
The tentative optical luminosity and efficiency of the source are similar to those of the Vela pulsar, 
which also supports the optical identification. 
However, the source spectrum is unusually steep, with a spectral index of about 5, 
which is not typical of optical pulsars.
The source offset from the radio position of PSR J1357$-$6429, which is in line with the corresponding 
offset of the X-ray position, 
implies the pulsar transverse velocity of 1600--2000 km s$^{-1}$ 
at the distance of 2--2.5 kpc, 
making it the fastest moving pulsar known. 
\end{abstract}

\section{Introduction}

\label{sec:int}

PSR J1357$-$6429 is a young (characteristic age $\tau$ = 7.3 kyr) 
and energetic (spin-down luminosity 
$\dot{E}$ = 3.1 $\times$ 10$^{36}$ ergs~s$^{-1}$) 
166 ms radio pulsar that was 
discovered in the Parkes multi-beam survey 
of the Galactic plane \citep{camilo2004ApJ}. 
At a distance of 2.4 kpc estimated from 
its dispersion measure (DM), 
it is one of the nearest young pulsars known. 
This proximity has motivated further  observations   
of the pulsar field in different spectral domains.  

The X-ray observations with \textit{Chandra}
and \textit{XMM-Newton}
have revealed an X-ray counterpart of the pulsar
and a tail-like   
pulsar wind nebula (PWN) \citep{esposito2007A&A,zavlin2007ApJ,chang2012ApJ,lemoine-goumard2011A&A}  
and have firmly established 
X-ray pulsations \citep{chang2012ApJ,lemoine-goumard2011A&A}. 
Periodic pulsations of PSR J1357$-$6429 were also discovered in the GeV range with \textit{Fermi} 
\citep{lemoine-goumard2011A&A}.
A fainter extended X-ray plerion was detected 
on a few tens of arcminutes scale \citep{chang2012ApJ,abramowski2011A&A}  
which positionally coincides with an extended TeV source, HESS J1356$-$645
\citep{abramowski2011A&A}. The TeV source is 
associated with the pulsar, and a radio plerion, a  supernova remnant
(SNR) candidate, catalogued as G309.8$-$2.6.

The age and observational properties of  the  J1357$-$6429 pulsar/PWN system 
appear to be  similar to the Vela pulsar system, 
which is about ten times closer to us and much more comprehensively 
studied in various spectral domains.   
In contrast to Vela, the J1357$-$6429 field has not been studied in the optical. 
To search  for  an optical counterpart  of  the  J1357$-$6429 pulsar/PWN system, we carried out  
the first deep observations of its field with the ESO Very Large Telescope (VLT) in $VRI$ bands. 
For details of observations and data reduction see \citet{danilenko2012A&A} while 
main results and conclusions are presented below. 

\begin{figure*}[t]
 \setlength{\unitlength}{1mm}
\begin{center}\includegraphics[scale=0.5, clip=]{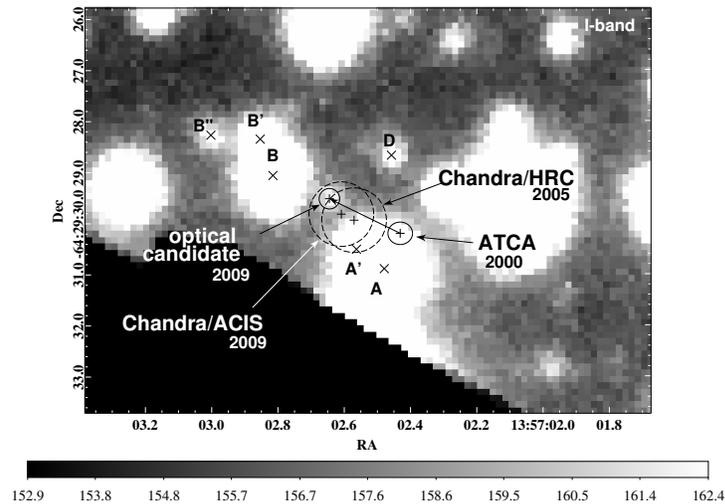}
\end{center}
 \caption{The fragment 
 of the VLT $I$-band image.  
 The 1$\sigma$ error ellipses of the X-ray and radio positions 
 of the pulsar are shown. 
 The position of the possible optical counterpart and its uncertainties 
 are also indicated along with possible pulsar proper-motion path.  
 Background sources located near/along this path are also marked.}
 \label{fig:1}
 \end{figure*}

\section{Selection of a counterpart candidate}
\label{sec:diss}
 \begin{figure*}[t]
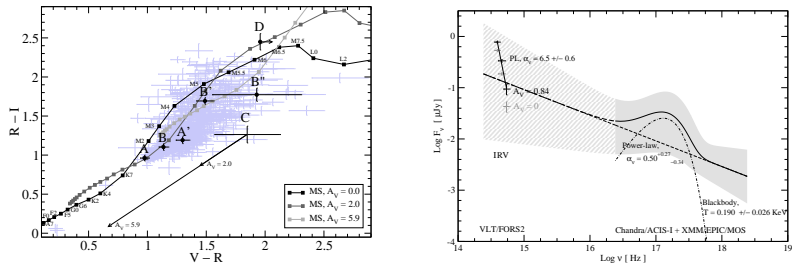

\setlength{\unitlength}{1mm}
\resizebox{11.5cm}{!}{
\begin{picture}(50,14)(1,0)
\put (5,0) {\includegraphics[scale=0.09,clip]{V-R_vs_R-I.bw.eps}}
\put (30,0) {\includegraphics[scale=0.09, clip]{mw_spectrum_1_reg.bw.eps}}
\end{picture}}
 \caption{{\sl Left:} The colour-colour 
diagram of PSR J1357$-$6429 field stars.
The candidate labelled by C and nearby stars marked in Fig.~\ref{fig:1} are shown. 
Empirical main-sequence curves reddened with a zero,  medium, and  
entire Galactic extinction, $A_V$ = 0, 2.0, and 5.9,
are shown. {\sl Right:} Multi-wavelength unabsorbed spectrum for PSR J1357$-$6429.  
 The best fits to the X-ray spectrum,
 including
 thermal and non-thermal components, 
 are shown along with the observed 
 and dereddened optical fluxes.}
 \label{fig:2}
 \end{figure*}
A fragment of the VLT $I$-band image presented in Fig.~\ref{fig:1} depicts 
vicinity of the pulsar. 
At least three, $A$, $A'$, and $C$, of the nearby objects marked in Fig.~\ref{fig:1}
can be considered as potential optical counterparts of the pulsar.
Object $A$ has been
already analysed and qualified as a mere main sequence 
star by \citet{mignani2011A&A} based on a part of our VLT data.  
To select a most plausible candidate to the optical counterpart of the pulsar
we performed the PSF photometry on our $VRI$ images and constructed
colour-colour and colour-magnitude diagrams and the former one is presented in 
Fig.~\ref{fig:2} while the rest can be found in \citet{danilenko2012A&A}.
Colours of the object $C$ differ from colours of the most part of the field stars
but compatible with ones of some other pulsar optical
counterparts usually detected as faint blue objects,
like, for instance, the Geminga-pulsar optical counterpart   
\citep{shibanov2006A&A}.

Besides that, among the other potential candidates, 
the spatial position of object $C$ 
is  most closely  compatible with the X-ray position of the pulsar obtained  from  
the ACIS-I observations taken in the same year as the optical data.
All the above allows us to 
keep it as the only likely  optical counterpart 
of the pulsar J1357$-$6429.

If it is indeed the counterpart, this implies 
at least two prominent properties of the pulsar.

\section{Unusually steep spectrum}
The multi-wavelength spectrum constructed from the $VRI$ fluxes of the counterpart and 
the X-ray data reanalysed by
us \citep{danilenko2012A&A} is in Fig.~\ref{fig:2}. The optical fluxes
were dereddened with extinction, A$_{V}$, of 0.8 based on hydrogen column 
density, $N_{H}$, estimated
by means of BB+PL fit to the X-ray data \citep{danilenko2012A&A}.
The putative counterpart has an extremely red spectrum which is atypical of pulsars and    
raises   doubts about the   pulsar nature of the candidate.  
Several possible explanations of the spectrum steepness were proposed by 
\citet{danilenko2012A&A}.

We also estimated the counterpart optical luminosity
(in the $V$ band) of 1.5 $\times$ 10$^{29}$ ergs s$^{-1}$, 
and efficiency, $\eta_{V}$ = $L_{V}$/$\dot{E}$, of 
4.8 $\times$ 10$^{-8}$, 
to be close to that of 
the Vela pulsar, which has similar age and is known to be 
inefficient in the optical and X-rays 
\citep{zharikov2006AdSpR}.   
The counterpart $L_{X}/L_{V}$ ratio of about 
100 is 
also in the range of 100--1000, typical of pulsars.  
All that are indirect evidence in favour of the real
identification of the counterpart candidate with the pulsar.

\section{Very high proper motion}
A significant offset of the counterpart  candidate position 
from the  J1357$-$6429 radio coordinates   
measured  8.7 yr earlier, 1\farcs54 $\pm$ 0\farcs32 (Fig.~\ref{fig:1}),   
suggests that the pulsar has a high  proper motion.       
At the most plausible distance range of 2--2.5 kpc, which 
follows from the A$_{V}$--distance relation and N$_{H}$ estimated from the X-ray data 
\citep{danilenko2012A&A}, this implies 
the pulsar transverse velocity to be between    
1690 $\pm$ 350 km s$^{-1}$ and 2110 $\pm$ 440 km s$^{-1}$.
The value is opposed to 
a two-to-three times smaller value 
based on a 7 arcmin offset of the pulsar from 
the centre of the extended source HESS J1356$-$645 \citep{abramowski2011A&A}.
But, the difference between two estimates can be reconciled by 
assuming that the true age of J1357$-$6429 is twice as small as the characteristic one,
and two ages may differ by at least 40\% \citep[see, e.g.][]{brisken2003AJ}. 
The direction of the putative proper motion is nearly consistent with the
direction of the 7 arcmin offset and with 
the  NE extension of the  tail-like PWN structure detected 
in X-rays \citep{chang2012ApJ, zavlin2007ApJ}.
Besides that, a significant difference between the radio interferometric  
\citep{camilo2004ApJ} and X-ray pulsar 
positions  obtained at different epochs is also indicative of a proper motion of the pulsar 
as noted by \citet{mignani2011A&A}.   
From Fig.~\ref{fig:1}, one can see that
the  \textit{Chandra}/ACIS position is  
apparently shifted from  the \textit{Chandra}/HRC one,  
in line with the suggested motion, although  
the shift is of a low significance.

To our knowledge, the highest  pulsar velocity of 1080 $\pm$ 100 km s$^{-1}$,  
which has been firmly established  
by direct proper-motion and parallax measurements with the VLBA,   
belongs to PSR B1508+55 \citep{chatterjee2005ApJ}. 
The distance to that pulsar, 2.37 $\pm$ 0.20 kpc,  
is comparable to that of J1357$-$6429, which suggests
that similar direct measurements   
are also possible for the latter. 
To confirm the high proper motion, it is necessary to repeat 
radio observations performed with the ATCA \citep{camilo2004ApJ} once again and then, if the high proper motion is
confirmed, to perform parallax measurements with the VLBA to confirm the 
distance to PSR J1357$-$6429. As a result we may establish the fastest pulsar 
ever known, which is important for understanding the nature of high  
velocities of pulsars \citep[see, e.g.][]{chatterjee2005ApJ}.  
These observations would  also be 
of a great complement to the approved VLT/NACO high spatial
resolution adaptive optical observations of the pulsar field and 
they can firmly confirm the
optical counterpart by comparing the proper motion measurements in the optical 
and radio ranges.

\acknowledgements The work was partially supported by the Russian Foundation for Basic 
Research (grants 11-02-00253 and 11-02-12082),
Rosnauka (Grant NSh 4035.2012.2), and 
the Ministry of  Education and Science of the Russian Federation (Contract No. 11.G34.31.0001).
SZ acknowledges support from CONACYT 151858  project,
GP was partly supported by NASA grant NNX09AC84G, and
REM acknowledges support by the BASAL
Centro de Astrof\'isica y Tecnologias Afines (CATA) PFB--06/2007.

\bibliography{ref-2.bib}
\end{document}